\documentstyle[preprint,tighten,aps,floats]{revtex}

\begin{document}
\draft
\title{ 
Third-harmonic exponent in three-dimensional $N$-vector models
}
\author{Martino De Prato,${}^{a}$ 
        Andrea Pelissetto,${}^{b}$
        Ettore Vicari${}^{c}$
}
\address{${}^a$ Dipartimento di Fisica, Universit\`a di Roma Tre, and 
              INFN, I-00146 Roma, Italy} 
\address{${}^b$ Dipartimento di Fisica, Universit\`a di Roma La Sapienza,
and INFN, I-00185 Roma, Italy} 
\address{${}^c$ Dipartimento di Fisica, Universit\`a di Pisa, and 
              INFN,
              I-56127 Pisa, Italy \\ 
{\bf e-mail: \rm 
{\tt deprato@fis.uniroma3.it},
{\tt Andrea.Pelissetto@roma1.infn.it},
{\tt vicari@df.unipi.it}.
}}

\maketitle

\begin{abstract}
We compute the crossover exponent associated with the spin-3 operator
in three-dimensional $O(N)$ models. A six-loop field-theoretical calculation 
in the fixed-dimension approach gives $\phi_3 = 0.601(10)$ 
for the experimentally relevant case $N=2$ (XY model). The corresponding 
exponent $\beta_3 = 1.413(10)$ is compared with the experimental estimates 
obtained in materials undergoing a normal-incommensurate structural
transition and in liquid crystals at the smectic-A--hexatic-B phase
transition, finding good agreement.
\end{abstract}


\newpage

In nature many physical systems undergo phase transitions belonging
to the universality classes of the  O($N$) vector models. In particular,
the XY model, corresponding to $N=2$, describes the $\lambda$-transition
in ${}^4$He, (anti)ferromagnets with easy-plane anisotropy,
density-wave systems, etc.; see Ref.~\cite{review} for a review. The critical
exponents associated with the order parameter have been accurately measured 
both experimentally and theoretically \cite{review}. Moreover, in some 
XY systems it is also possible to measure experimentally the critical exponents
associated with secondary order parameters. This is the case of liquid
crystals 
\cite{Brock-etal-86,ABBL-86,Aharony-etal-95},
of normal-incommensurate transitions
\cite{AM-83,HHTG-95,ZMHCLGGS-96}, and of graphite-intercalation compounds 
\cite{Bak-80}. 

The most relevant exponent is the second-harmonic one that has been
recently computed to high-precision using field-theoretical methods
in Refs.~\cite{CPV-l2,CPV-l4}. The $\epsilon$-expansion gives 
$\phi_2 = 1.174(12)$, while the fixed-dimension expansion gives
$\phi_2 = 1.184(12)$. The fourth-harmonic crossover exponent 
was reported in Refs.~\cite{CPV-00,CPV-l4}: 
$\phi_4 = -0.077(3)$ ($\epsilon$ expansion)
and $\phi_4 = -0.069(5)$ (fixed-dimension expansion).
Here, we wish to determine 
the third-harmonic exponent by means of a six-loop perturbative calculation in 
the fixed-dimension approach, extending previous three-loop determinations 
\cite{WH-74,Wallace-76,ABBL-86}. 
Such a calculation is also relevant for some crossover phenomena,
in which the XY symmetry is reduced to that of the three-state Potts 
model, as it happens in cubic magnets in the presence of stress or 
of appropriate magnetic fields \cite{AMB-77,MFD-76}.

In the field-theoretical approach one starts from the usual 
$\phi^4$ Hamiltonian
\begin{equation}
{\cal H}= \int d^dx\left[ 
   {1\over 2}(\partial_\mu {\phi})^2
 + {1\over 2} r {\phi}^2 + {1\over 4!}u ({\phi}^2 )^2\right],
\label{Hphi4}
\end{equation}
where ${\phi}_a(x)$ is an $N$-component real field. 
The XY model corresponds to $N=2$, but here we will keep $N$ generic.
Secondary order parameters are associated with operators 
${\cal O}^{(l)}$ that are polynomials of order $l$ in the fields
and that transform irreducibly under the spin-$l$ representation of the 
$O(N)$ group. In particular, the third-harmonic operator is 
\begin{equation}
{\cal O}^{(3)}_{abc} = 
  \phi_a \phi_b \phi_c - {\phi^2\over N + 2} 
   \left(\phi_a \delta_{bc} + \phi_b \delta_{ac} + \phi_c \delta_{ab}\right).
\end{equation}
We wish now to compute the crossover exponent $\phi_3$ associated to 
${\cal O}^{(3)}_{abc}$ and the corresponding exponents $\beta_3$ and 
$\gamma_3$ given by
\begin{eqnarray}
\beta_3 &=& 2 - \alpha  - \phi_3,
\nonumber \\
\gamma_3 &=& - 2 + \alpha + 2\phi_3.
\label{scalrel}
\end{eqnarray}
The exponents $\beta_3$ and  $\gamma_3$ describe
respectively the critical (singular) behavior of
the average $\langle {\cal O}^{(3)}(x) \rangle \sim |t|^{\beta_3}$ and
of the susceptibility
$\chi_{\cal O} \equiv 
    \sum_x \langle {\cal O}^{(3)}(0) {\cal O}^{(3)}(x) \rangle_c
    \sim |t|^{-\gamma_3}$.  

For this purpose we determine  the
renormalization function $Z_3(g)$ from the one-particle irreducible
three-point function $\Gamma_3^{(3)}(0)$ with an insertion of the operator
${\cal O}^{(3)}_{abc}$ at zero external momenta, i.e. we set
\begin{equation}
\langle {\cal O}^{(3)}_{abc} \phi_a \phi_b \phi_c \rangle^{1PI} = 
   A Z_3^{-1}(g),
\end{equation}
where $A$ is a numerical coefficient that ensures that 
$Z_3(0)=1$, and $g$ is the four-point renormalized coupling.
Then, we compute  the renormalization-group function 
\begin{equation}
\eta_3(g) \equiv \left. {\partial \ln Z_3 \over \partial \ln m} \right|_u 
= \beta(g) {d \ln Z_3 \over d g}, 
\label{defesponenteeta3}
\end{equation}
and $\eta_3 = \eta_3(g^*)$, where $g^*$ is the fixed-point value of $g$.
Finally, the renormalization-group scaling relation (valid in three 
dimensions)
\begin{equation}
\label{scalrel2}
\phi_3=\left(\eta_3+\frac{3}{2}-\frac{3}{2}\eta\right)\nu
\end{equation}
allows us to determine $\phi_3$.

We computed $\Gamma^{(3)}_3(0)$ to six loops. 
The calculation is rather cumbersome since it requires
the evaluation of a few thousand Feynman  diagrams.
We handled it with a symbolic manipulation program, 
which  generates the diagrams 
and computes the symmetry and group factors of each of them.
We used the numerical results compiled in Ref.~\cite{NMB-77}
for the integrals associated with each diagram.
We obtained
\begin{eqnarray}
\eta_3(\bar{g}) &=& -{6\over N+8} \bar{g} + {2 (N+10)\over (N+8)^2} \bar{g}^2
   -  {128.736 +15.4900\,N - 0.650238\,{N^2} \over (N+8)^3} \bar{g}^3
\nonumber \\
 && + {1148.68 + 191.005\,N + 1.82163\,{N^2} + 0.283028\,{N^3} \over (N+8)^4}
      \bar{g}^4
\nonumber \\ 
 && - {12606.9 + 2550.46\,N + 64.4818\,{N^2} - 2.34060\,{N^3} - 0.152501\,{N^4}
     \over (N+8)^5} \bar{g}^5 
\nonumber \\
 && + {161373. + 38736.8\,N + 1874.23\,{N^2} - 5.98451\,{N^3} + 1.88168\,{N^4} 
      + 0.094179\,{N^5}\over (N+8)^6} \bar{g}^6 
\nonumber \\ 
 && + O(\bar{g}^7), 
\end{eqnarray}
where, as usual, we have introduced the rescaled coupling $\bar{g}$
defined by 
\begin{equation}\label{gnew}
g  =  {48 \pi\over 8+N} \;\bar{g} .
\end{equation}
Here $g$ is the usual four-point renormalized coupling normalized so that 
$g = u/m$ ($m$ is the renormalized mass) at tree level.

Field-theoretical  perturbative expansions are divergent, and thus,
in order to obtain accurate results, an appropriate resummation 
is required. We use the method of Ref.~\cite{LZ-77} that takes into account 
the large-order behavior of the perturbative expansion, 
see, e.g., Ref.~\cite{Zinn-Justin-book}. Mean values and error
bars are computed using the algorithm of Ref. \cite{CPV-00}.

Given the expansion of $\eta_3(\bar{g})$, 
we determine the perturbative expansion of 
$\phi_3(\bar{g})$, $\beta_3(\bar{g})$, and 
$\gamma_3(\bar{g})$ using relations (\ref{scalrel}) and 
(\ref{scalrel2}). Then, we resum the perturbative series and compute
them at $\bar{g} = \bar{g}^*$ \cite{foot1}.
For $N=2$ we obtain $\phi_3=0.5963(21)$, $0.5968(2)$, 
                    $\beta_3=1.398(8)$, $1.405(3)$, 
                    $\gamma_3=-0.800(7)$, $-0.808(13)$, 
where for each exponent we report the estimate obtained from the direct 
analysis and from the analysis of the series of the inverse. 
The two estimates obtained for each exponent agree within error bars, but, 
with the quoted errors, the scaling relations (\ref{scalrel}) are not well 
satisfied. For instance, using $\nu=0.67155(27)$ (Ref. \cite{CHPRV-01}) 
and $\beta_3=1.403(8)$ we obtain $\phi_3=0.611(8)$, 
while using the same value of $\nu$ and $\gamma_3=-0.803(13)$ 
we have $\phi_3=0.606(6)$. These two estimates are slightly higher than 
those obtained from the analysis of $\phi_3(g)$ and $1/\phi_3(g)$. 
Clearly, the errors are somewhat underestimated, a phenomenon
that is probably connected with the nonanalyticity
\cite{PV-98,CCCPV-00,CPV-01} of the
renormalization-group functions at the fixed point $\bar{g}^*$. 

In order to obtain a conservative estimate, we have thus decided to take 
as estimate of $\phi_3$ the weighted average of the direct estimates and of 
the estimates obtained using $\beta_3$ and $\gamma_3$ together with the 
scaling relations (\ref{scalrel}). 
The error is such to include all estimates. 
The other exponents are dealt with analogously \cite{foot2}. 
The final results for several values of $N$ are reported in 
Table~\ref{esponenti}.
 
\begin{table}[!tbp]
\caption{Critical exponents associated with the spin-3 
operator $O_{abc}^{(3)}$}
\label{esponenti}
\begin{tabular}{cccc}
$N$ & $\phi_3$ & $\beta_3$ & $\gamma_3$  \\
\hline
0 & 0.445(11) &  1.331(11) & $-$0.89(2) \\
2 & 0.601(10) &  1.413(10) & $-$0.81(2) \\
3 & 0.678(18) &  1.455(18) & $-$0.78(4) \\
4 & 0.760(23) &  1.487(23) & $-$0.73(4)   \\
5 & 0.814(14) &  1.484(14) & $-$0.67(3)   \\
8 & 0.971(33) &  1.519(33) & $-$0.55(7)   \\
16 & 1.193(12)&  1.540(12) & $-$0.35(2)   \\
$\infty$ & ${3\over2}$ & ${3\over2}$ & 0 \\
\end{tabular}
\end{table}

Let us compare these results with previous ones for $N=2$. 
Ref.~\cite{ABBL-86} reports $\beta_3 = 3 \beta + 3 \nu x_3$, 
where $x_3 \approx 0.174$ from the $\epsilon$-expansion and 
$x_3 \approx 0.276$ from the fixed-dimension expansion. 
It follows $\beta_3 \approx 1.397$ and $\beta_3 \approx 1.602$ in the two cases.
These results are reasonably close to ours. The exponent $\beta_3$ 
has been determined at the smectic-A--hexatic-B phase transition 
in liquid crystals, obtaining \cite{Brock-etal-86,ABBL-86}
$\beta_3\approx 4.8\beta \approx 1.66$,
using $\beta = 0.3485(4)$ (Ref.~\cite{CHPRV-01}). Such an exponent has also 
been measured in some materials exhibiting a structural normal-incommensurate
phase transition. From the analysis of x-ray scattering data in erbium
Ref.~\cite{HHTG-95} obtains $\beta_3 = 1.8(3)$,
while two different experiments in Rb$_2$ZnCl$_4$ give
respectively $\beta_3 = 1.80(5)$ 
(Ref.~\cite{AM-83}) and $\beta_3 = 1.50(4)$
(Ref.~\cite{ZMHCLGGS-96}). Keeping into account that the experimental 
errors seems to be underestimated, there is  
reasonable agreement with our results. The exponent 
$\delta^* \equiv \beta/\phi_3$ was measured at the trigonal-to-pseudotetragonal
transition in [111]-stressed SrTiO$_3$ obtaining \cite{AMB-77} 
$\delta^*=0.62(10)$. 
Such an estimate is in good agreement with our prediction 
$\delta^* = 0.58(1)$.  Finally, we should mention that 
our results are also relevant for polymer physics. Indeed, we can derive from
the estimates obtained for $N=0$ the partition-function exponent 
$p$ for nonuniform \cite{foot3} star polymers with three arms: 
$p = 3(\gamma + \nu)/2 + \phi_3 = 3.06(1)$, where we used 
$\gamma = 1.1575(6)$ (Ref.~\cite{CCP-98}).


\end{document}